# Impacts of generative AI skill development on academic adaptation and learning outcomes among university students

Lei Fan[*,1] and Fangxue Liu[1,2]

[1] Department of Civil Engineering, Xi'an Jiaotong-Liverpool University, Suzhou, 215123, China
[2] Department of Computer Science, University of Liverpool, Liverpool, L69 3BX, UK
*Corresponding Author (Email: Lei.Fan@xjtlu.edu.cn)

**Abstract**: Understanding how university students perceive the development of their generative artificial intelligence (GenAI) skills over time can clarify how GenAI use relates to learning and academic outcomes. This study examined perceived skill-development trajectories, changes in learning practices, and associated academic gains using a retrospective questionnaire with 313 analyzed responses and interviews with 18 students across subject areas. Descriptive and comparative analyses and exploratory structural equation modeling (SEM) were integrated with thematic analysis. Perceived developmental trajectories were heterogeneous: 34.2% of students reported rapid early improvement followed by a plateau, 28.4% steady improvement, and 21.1% delayed acceleration. Development was also uneven across skills, with improvement in evaluating GenAI outputs perceived as more challenging; assignments were the most frequently selected driver of skill growth. Interviews further indicated increasingly specific prompting, tailored verification, and selective delegation, although some students reported stable routines, dependence, skipped reasoning, or improvements attributable partly to better GenAI tools. Academic gains were reported more often for assignments than for examinations. The SEM indicated a positive indirect association between perceived skill growth and academic gains through adaptation of assignment processes, whereas academic gains did not clearly differ across developmental trajectory groups after multiplicity correction. The findings suggest that perceived GenAI skill development varies in both trajectory and skill domain and is associated with changes in learning practices and differentiated academic benefits.

## 1. Introduction

As university students continue to use generative artificial intelligence (GenAI), a central educational question is how their GenAI skills change beyond initial familiarity. Students may learn to frame more precise requests, recognize unreliable answers, and decide which parts of academic work require their own reasoning (Yan et al., 2025). These changes need not occur together or continue at a constant pace. Understanding students' perceived development therefore requires attention to what they believe has changed, how they describe changes over time, and whether these changes are associated with different ways of learning and performing academic tasks. Artificial intelligence (AI) literacy frameworks support this broader view by distinguishing the application of AI from its evaluation and ethical use (Long & Magerko, 2020; Ng et al., 2021).

The academic significance of developing GenAI skills also requires a distinction between producing better work with GenAI assistance and developing learning that can be demonstrated independently. In a university writing experiment, Y. Fan et al. found greater improvement in essay scores with ChatGPT support but no statistically significant advantage in knowledge gain or transfer (Y. Fan et al., 2025). Ma et al. reported opposing indirect associations between GenAI assistance and academic performance through self-efficacy and thinking offloading (Ma et al., 2026). These findings make changes in learning behaviour central to a developmental account: increasingly fluent use may support explanation, feedback, and revision, but may also make it easier to delegate reasoning.

Repeated-observation studies have begun to trace how GenAI practices change over time. Aruleba et al. followed 21 computer-science undergraduates at one UK university for four weeks, documenting more specific prompting and critical engagement alongside continued passive reliance among some students (Aruleba et al., 2025). Tan's year-long case study traced a graduate student's expanding use of GenAI within disciplinary learning and academic participation, while Luo and Yin examined uneven engagement in 11 students' assessment-related dialogues over a semester (Tan, 2025; Luo & Yin, 2026). At a broader scale, Polyportis followed 222 students across Dutch higher-education institutions over eight months, examining changes in ChatGPT use and factors associated with those changes (Polyportis, 2024). These studies establish the value of repeated observation for understanding changing strategies, purposes, and engagement.

Their scope nevertheless leaves room for complementary evidence. Intensive developmental studies offer temporal detail but often involve small, discipline-specific or single-institution samples, limiting coverage of diverse educational experiences (Aruleba et al., 2025; Tan, 2025). Comparatively larger samples can broaden participation while focusing on a narrower set of adoption-related variables, rather than connecting perceived skill development with changes in learning practices and academic outcomes (Polyportis, 2024). Existing student surveys also describe benefits and educational expectations, but leave these developmental dimensions insufficiently connected (Chan & Hu, 2023; L. Fan et al., 2025). Examining them can show how students' perceived developmental patterns relate to specific skill improvements, changes in academic practices, and reported academic gains, complementing the depth of repeated-observation studies.

This study addresses that gap using a retrospective questionnaire with 313 analyzed responses and interviews with 18 university students across academic disciplines. It brings together perceived developmental trajectories, specific skill improvements, changes in assignment processes, reported academic gains, and support demands. Recruitment across diverse disciplines and geographic regions, including science, technology, engineering, and mathematics (STEM) and humanities and social sciences (HSS), extends the evidence beyond narrowly bounded disciplinary settings in repeated-observation studies.

This study contributes an integrated account of the heterogeneity in students' perceived GenAI skill development and its association with learning practices and reported academic outcomes. By identifying variation in developmental trajectories, learning adaptations, and academic benefits, the study also highlights patterns that warrant further investigation, particularly through longitudinal research capable of measuring actual change over time. Accordingly, three research questions guide the study.

- What specific skill improvements and retrospectively perceived developmental trajectories do university students report in their academic use of GenAI?
- How are perceived GenAI skill growth, assignment-process adaptation, and academic gains associated with one another and with reported developmental trajectories, student characteristics, and GenAI-use patterns?
- How do students perceive learning support and assessment alignment, and what institutional support do they recommend?

## 2. Literature review

### 2.1. Temporal development of GenAI skills

Development in academic GenAI use involves changes in operational skill, evaluative judgment, and the ability to integrate assistance into purposeful work (Yan et al., 2025; Walton et al., 2026). Long and Magerko's competency framework and Ng et al.'s conceptualization of AI literacy position effective use within a broader capability set than tool operation alone (Long & Magerko, 2020; Ng et al., 2021). However, these frameworks identify dimensions of competence rather than a universal developmental sequence. This distinction is important because confidence, frequency of use, and competence represent different outcomes. Hornberger et al.'s AI-literacy test assesses knowledge, while O'Dea et al.'s survey found lower perceived confidence in evaluation and creation than in understanding and application (Hornberger et al., 2023; O'Dea et al., 2026). Neither knowledge nor confidence at a single time point demonstrates development.

Temporal studies also differ in what they measure. Polyportis' two-wave study of 222 students found declining reported ChatGPT use over eight months, with changes associated with adoption-related perceptions (Polyportis, 2024). This does not indicate whether reduced use reflects greater selectivity or declining competence. Williams compared successive first-year medical-science cohorts and found broader use and greater confidence in the later cohort, reflecting changing cohort contexts rather than individual development (Williams, 2026). Together, these studies show that changes in adoption and capability require separate explanations rather than a single developmental trajectory.

Studies following academic activity provide more direct evidence of changing practice. Aruleba et al. tracked 21 computer-science undergraduates through structured activities, journals, and reflection over four weeks. Many developed more specific requests, iterative interactions, and greater scrutiny of generated code, although some continued to accept or copy solutions (Aruleba et al., 2025). Tan followed one graduate student over an academic year and described GenAI becoming integrated into interpreting disciplinary expectations, understanding genres, and developing research ideas, alongside concerns about identity and learning losses (Tan, 2025). These studies suggest that development concerns the relationship between GenAI use and academic demands, but their limited settings do not establish common stages or rates of progression.

### 2.2. Changing learning practices and critical engagement

A developmental account must consider how GenAI changes participation in academic work. Self-regulated learning links goal setting and strategy selection with monitoring and reflection (Panadero, 2017; Zimmerman, 2002). Applied to GenAI, this perspective highlights whether students formulate problems, evaluate responses against evidence, revise approaches, and judge their own understanding. Increased interaction alone cannot demonstrate these processes; development is educationally meaningful when prompting changes the quality of participation rather than merely the speed of obtaining answers.

Luo and Yin analysed authentic GenAI dialogues from 11 university students completing assessed work over a semester. They identified behavioural, cognitive and metacognitive, emotional, and agentic engagement, with 12% of analysed dialogues showing agentic engagement, including challenging outputs and directing assistance according to assessment needs (Luo & Yin, 2026). These forms of engagement should not be interpreted as developmental stages. Combined with Aruleba et al.'s findings of increasingly specific but unevenly critical interactions, the evidence suggests that sustained access can coexist with limited active direction of GenAI assistance (Aruleba et al., 2025). Fluent requesting therefore remains distinct from the judgment required to question or reject GenAI outputs.

Verification and control of delegation are central to this distinction. Students may refine prompts until responses appear suitable without establishing their validity, whereas deliberate uses such as explanation or critique may preserve learner involvement. AI-literacy frameworks emphasise evaluation and ethical responsibility, while academic-literacy research situates AI decisions within disciplinary participation and identity (Ng et al., 2021; Tan, 2025). Development may therefore involve changing boundaries around GenAI use as well as improved operational skills. Whether such changes sustain independent reasoning requires examining tasks, decisions, and learning processes rather than usage behaviour alone.

### 2.3. Academic consequences of developing GenAI skills

The relationship between GenAI skill development and academic outcomes depends on the outcome assessed. Assisted assignment quality, subject understanding, and performance without assistance may respond differently to the same tool. Y. Fan et al.'s randomized study of 117 university students illustrates this distinction: the ChatGPT group achieved greater essay-score improvement, but knowledge gain and transfer did not significantly differ across support conditions (Y. Fan et al., 2025). Thus, improved assisted performance cannot automatically be interpreted as improved independent capability.

Instructional design further shapes the consequences of assistance. Kestin et al. found substantial learning benefits from a purpose-built GenAI tutor in undergraduate physics (Kestin et al., 2025). In high-school mathematics, Bastani et al. reported improved GenAI-supported practice but poorer subsequent unaided examination performance with a standard-style interface; instructional safeguards reduced this effect without demonstrating a significant unaided-performance advantage (Bastani et al., 2025). These findings concern specific interventions rather than natural skill development, but they demonstrate why GenAI design and usage conditions matter when interpreting academic outcomes.

Broader evidence supports a conditional interpretation. Gu and Yan's meta-analysis of 19 studies found a positive average effect of GenAI interventions on academic performance, with substantial heterogeneity across studies (Gu & Yan, 2025). In a survey of 443 Chinese university students, Ma et al. identified positive and negative indirect statistical associations through self-efficacy and thinking offloading, respectively (Ma et al., 2026). Although such associations do not establish causal mechanisms, they highlight competing pathways through which GenAI assistance may influence learning. Understanding development therefore requires examining whether growing GenAI skill corresponds to changed learning practices and whether reported academic benefits extend beyond immediate assisted tasks.

## 3. Methodology

### 3.1. Study design and data collection

The study used a mixed-methods design, combining a cross-sectional retrospective questionnaire with interviews to examine university students' perceived GenAI skill development and associated academic experiences. The questionnaire was administered online through Wenjuanxing from early July to mid-August 2026. Convenience sampling was combined with targeted recruitment to broaden participation across gender, academic disciplines, and geographic regions, covering different parts of China. Of over 400 questionnaire responses collected, 313 were analyzed; the remaining responses were excluded based on data-quality screening for short completion times, duplicate submissions, or identical responses across all agreement items. All quantitative analyses used this final dataset.

Eighteen students (named P$i$ where $i$ = 1 – 18) with experience using GenAI were selected for interviews from diverse academic disciplines. Interviews took place in late July and early August 2026, and lasted approximately one hour each. All interviewees also participated in the original questionnaire, but survey anonymity prevented individual records from being linked in the questionnaire sample. Both components relied on retrospective self-report and did not include repeated competency measurements or independently verified performance changes.

The study protocol was approved by the University Research Ethics Review Panel of Xi'an Jiaotong-Liverpool University on 13 Jun 2026. The survey and interviews were conducted in accordance with applicable ethical guidelines and regulations, and informed consent was obtained from all participants prior to their participation.

### 3.2. Questionnaire structure

The questionnaire comprised 39 response fields, including 26 agreement items and 13 background, experience, behavioral, and open-response fields (Table 1). The agreement items used a five-point scale from 1 (strongly disagree) to 5 (strongly agree). The perceived-growth items refer retrospectively to change since initial academic GenAI use, and the process and academic-gains items explicitly attribute changes to developing GenAI skills. Other agreement blocks assess perceived sources of development, current views of assessment, or responsible-use practices. Appendix A provides all survey items.

### 3.3. Interview scope and analysis

The interviews explored participants' initial experiences with GenAI, perceived stages and turning points in GenAI skill development, evaluation of GenAI outputs, sources of support, recent assignment use, delegation and dependence, and perceived academic gains. Participants represented a range of disciplines, including business, finance, communication, language studies, biological sciences, horticulture, mathematics, engineering, and computing.

Interview recordings were organized into a participant-by-topic matrix containing relevant excerpts and analytical summaries. The interview data were used to contextualize and qualify the survey findings by identifying areas of convergence, elaboration, and divergence across the two data sources.

### 3.4. Sample characteristics

The questionnaire sample characteristics are shown in Table 2, which included both undergraduate and postgraduate students, with gender representation approximately balanced. Disciplines were broadly classified as STEM or HSS; these categories were used only as contextual descriptors rather than for detailed disciplinary comparisons. In terms of GenAI experience, the largest group reported having used GenAI for academic purposes for 1–2 years.

Table 1. Questionnaire structure and analytical domains.

| Domain | Item codes | Item numbers |
|---|---|---|
| Background and experience | A1–A7 | 7 |
| Perceived growth | B1–B4 | 4 |
| Practice-related development | B5–B8 | 4 |
| Sources of support | B9–B12 | 4 |
| Developmental trajectory | B13 | 1 |
| Assignment-process adaptation | C1–C4 | 4 |
| Perceived academic gains | D1–D3 | 3 |
| Task-specific gains | E1 | 1 |
| Assessment alignment | F1–F3 | 3 |
| Responsible use | G1–G4 | 4 |
| Use intensity | H1–H3 | 3 |
| University support suggestions (open -response) | I1 | 1 |

Table 2. Sample characteristics and GenAI experience. Five respondents provided insufficient discipline information and were retained as unclassified; therefore, STEM and HSS percentages sum to 98.4% of the full sample ($n$ = 313). Minor rounding differences may occur.

| Characteristic | Category | $n$ | % |
|---|---|---|---|
| Gender | Female | 158 | 50.5 |
| | Male | 155 | 49.5 |
| Academic level | Undergraduate | 188 | 60.1 |
| | Postgraduate | 125 | 39.9 |
| Discipline | STEM | 134 | 42.8 |
| | HSS | 174 | 55.6 |
| Time since first academic use | Less than 6 months | 81 | 25.9 |
| | 6–12 months | 46 | 14.7 |
| | 1–2 years | 112 | 35.8 |
| | 2–3 years | 55 | 17.6 |
| | More than 3 years | 19 | 6.1 |

### 3.5 Descriptive and comparative analyses

Categorical responses were summarized using counts and percentages, while scale items and composites were described using means and standard deviations. For agreement items, endorsement was defined as responses of "agree" or "strongly agree". Six composite scores were constructed and used in the descriptive, trajectory, and correlation analyses: perceived skill growth (B1 − B4), practice-related development (B5 − B8), perceived assignment-process adaptation (C1 − C4), perceived academic gains (D1 − D3), perceived assessment alignment (F1 − F3), and responsible academic use (G1 − G4).

Spearman's rank correlation (ρ) was used to assess bivariate monotonic associations among composites and six contextual indicators: self-rated academic performance relative to peers (A4), time since first academic GenAI use (A5), self-rated digital literacy (A7), use frequency (H1), proportion of assignment tasks receiving GenAI assistance (H2), and interactions per assignment (H3). Spearman's ρ ranges from −1 to +1, where the sign indicates the direction of the association and the absolute value reflects the strength of the rank-based relationship (Schober et al., 2018).

Six within-person contrasts were examined: efficiency versus evaluative growth (B3 vs. B2), assignment versus examination gains (D1 vs. D2), online support versus university teaching (B10 vs. B11), peer support versus university teaching (B9 vs. B11), and assessment adaptation versus reflection (F3 vs. F1) and appropriateness (F3 vs. F2) (Table 5). Paired ordinal differences were tested using two-sided Wilcoxon signed-rank tests (Woolson, 2007), excluding zero differences and assigning average ranks to ties; *p*-values were based on the normal approximation without continuity correction. Mean paired differences were calculated using all pairs, including zero differences, with 95% percentile confidence intervals (CIs) obtained from 5,000 respondent-level bootstrap resamples that preserved paired observations. Standardized paired differences ($d_z$) were calculated as the mean difference divided by the standard deviation of individual differences. Because $d_z$ and the bootstrap confidence interval (CI) describe mean differences rather than rank differences, they complement rather than represent effect sizes for the Wilcoxon tests.

Tie-corrected Kruskal–Wallis tests (Kruskal & Wallis, 1952) compared outcome distributions across the five interpretable nominal trajectory groups (B13). The omnibus $H$ statistic was evaluated against a $\chi^2$ distribution with four degrees of freedom and indicates overall group differences rather than specific pairwise contrasts. Effect size was estimated using epsilon-squared, $\varepsilon^2 = \max[0, (H - k + 1)/(n - k)]$, with $k = 5$ and $n = 311$ (after excluding two unspecified "other" responses). Group means were accompanied by pointwise 95% percentile CIs from 5,000 within-group bootstrap resamples.

Holm correction (Holm, 1979) controlled familywise error across the six paired Wilcoxon tests, while Benjamini–Hochberg (BH) correction (Benjamini & Hochberg, 1995) was applied separately to the eight trajectory tests (six composites, time since first use (A5), and use frequency (H1)) and the 66 unique contextual correlations among the 12 variables (six composites and six contextual indicators). Adjusted values are reported as "Holm-adjusted *p*" or "BH-adjusted *p*," using 0.05 as the reference threshold. The 95% confidence intervals were not multiplicity-adjusted. Therefore, exclusion of zero does not necessarily correspond to an adjusted $p < 0.05$. Structural-model *p*-values were excluded from these correction families and are reported as exploratory, unadjusted robust-test *p*-values.

### 3.6 Measurement assessment

Measurement quality was assessed using complementary reliability, factorability, validity, and factor-model diagnostics. Internal consistency was summarized using Cronbach's $\alpha$ with bootstrap 95% confidence intervals. Factorability was examined using the Kaiser–Meyer–Olkin (KMO) statistic and Bartlett's test (Kaiser & Rice, 1974; Tobias & Carlson, 1969). Composite reliability (CR) and average variance extracted (AVE) were derived from separate ordinal one-factor models for each of the six construct blocks using standardized loadings ($\lambda$) and residual variances ($\theta$). These indices were interpreted jointly as evidence of within-construct measurement quality rather than definitive tests of construct validity.

Candidate measurement models were estimated for ordered responses using a two-step polychoric diagonally weighted least squares (DWLS) procedure (Flora & Curran, 2004; Li, 2016). Model fit was evaluated using the comparative fit index (CFI), root mean square error of approximation (RMSEA), and standardized root mean square residual (SRMR), Higher CFI and lower RMSEA/SRMR indicate closer approximate fit, but favorable fit alone is insufficient. Therefore, model assessment also examined convergence, standardized loadings, residual variances, and the latent correlation matrix. Solutions with negative residual variances or a non-positive-semidefinite latent correlation matrix were considered inadmissible. Construct distinctness was further examined using latent correlations and heterotrait–monotrait (HTMT) ratios (Henseler et al., 2015), with values approaching one indicating weak distinctness. All candidate models used the same sample ($n = 313$). Because the models were compared within the same sample without independent validation or scaled nested-model difference tests, their fit was interpreted as descriptive evidence of model plausibility rather than confirmatory validation.

### 3.7 Structural associations

A focused structural equation model (SEM) examined associations among perceived skill growth (B1 − B4), perceived assignment-process adaptation (C1 − C4), and perceived academic gains (D1 − D3). Practice-related development (B5 − B8) was excluded because of its substantial empirical overlap with perceived skill growth in the measurement analysis. The SEM estimated growth-to-process, process-to-gains conditional on growth, and growth-to-gains conditional on process paths. The indirect association was calculated as the product of the first two standardized coefficients, and the total association as the sum of the direct and indirect associations. Contextual indicators were not included as covariates.

Fully standardized coefficients ($\beta$) represent outcome standard-deviation changes per predictor standard deviation, conditional on other predictors in the same equation. Because the structural component was saturated, the estimated paths were interpreted as associations rather than evidence of causal ordering or mediation (Maxwell & Cole, 2007).

Uncertainty was assessed using case bootstrapping of the $n = 313$ respondents. Of 2,000 bootstrap attempts, 1,998 were retained for the 95% percentile CIs; one replication produced an inadmissible solution, and one was excluded because it met the prespecified near-boundary criterion. Reported $p$-values were based on sandwich delta-method standard errors and therefore need not correspond exactly to the percentile bootstrap CIs.

## 4. Results

### 4.1 Reliability, validity and model selection

The six candidate composites (defined in Section 3.5) had alpha estimates of 0.712 – 0.804, blockwise Pearson-based KMO values of 0.701 – 0.796, CR values of 0.794 – 0.863, and AVE values of 0.501 – 0.668 (Table 3). Responsible use had the highest descriptive mean, 4.25, and assessment alignment the lowest, 3.52. All six blocks had alpha, KMO, and CR point estimates above 0.70 and AVE above 0.50, although practice and process AVE only narrowly exceeded 0.50 and their alpha bootstrap CIs extended below 0.70. The overall Pearson-based KMO for the 22 construct items was 0.899, and every blockwise Bartlett test had $p < 0.001$. These statistics support the adequacy of the item correlation structure for factor analysis and provide preliminary evidence of internal consistency.

Several factor models were compared to examine construct distinctness, considering both approximate fit and model admissibility (Table 4). The seven-factor model including support sources showed favorable fit but was inadmissible because the latent correlation matrix had a negative minimum eigenvalue (−0.0071). The admissible six-factor, 22-item model also fit well (CFI = 0.979, RMSEA = 0.056, SRMR = 0.060), but growth and practice were weakly differentiated, with a latent correlation of 0.954 and an HTMT ratio of 0.964. Merging these constructs in an exploratory five-factor model produced similarly good fit (CFI = 0.978, RMSEA = 0.056, SRMR = 0.062). By contrast, a one-factor model fit substantially worse (CFI =

0.885, RMSEA = 0.125, SRMR = 0.102), arguing against a single-factor representation, although this does not rule out common-method bias (Podsakoff et al., 2003).

The focused three-factor model of perceived growth, assignment-process adaptation, and perceived academic gains was admissible, with standardized loadings of 0.483–0.891. Alternative models merging growth with process or process with gains showed poorer approximate fit, although no scaled nested-model difference tests were conducted. Given the study focus and the substantial overlap between growth and practice, the three-factor model was adopted as the focused exploratory measurement model for subsequent structural analysis.

Table 3. Composite descriptions and blockwise measurement evidence. M and SD describe equal-weight composite means and standard deviation. $\alpha$[CI] used 2,000 respondent resamples.

| **Construct** | **Items** | **M (SD)** | ***α* [95% CI]** | **KMO** | **CR** | **AVE** |
|---|---|---|---|---|---|---|
| Perceived growth | B1–B4 | 3.98 (0.60) | 0.778 [0.710, 0.827] | 0.787 | 0.826 | 0.544 |
| Practice-related development | B5–B8 | 3.99 (0.58) | 0.712 [0.623, 0.779] | 0.747 | 0.800 | 0.501 |
| Process adaptation | C1–C4 | 3.98 (0.60) | 0.722 [0.633, 0.785] | 0.740 | 0.794 | 0.501 |
| Perceived academic gains | D1–D3 | 3.83 (0.67) | 0.804 [0.742, 0.851] | 0.710 | 0.858 | 0.668 |
| Assessment alignment | F1–F3 | 3.52 (0.74) | 0.780 [0.724, 0.824] | 0.701 | 0.832 | 0.623 |
| Responsible use | G1–G4 | 4.25 (0.60) | 0.802 [0.752, 0.841] | 0.796 | 0.863 | 0.612 |

Table 4. Exploratory ordinal measurement-model comparisons.

| **Model** | ***N*** | **df** | **CFI** | **RMSEA** | **SRMR** | **Admissible** |
|---|---|---|---|---|---|---|
| Six factors, 22 items | 313 | 194 | 0.979 | 0.056 | 0.060 | Yes |
| Seven factors, 26 items | 313 | 278 | 0.972 | 0.060 | 0.064 | No |
| One factor, 22 items | 313 | 209 | 0.885 | 0.125 | 0.102 | Yes |
| Five factors, merged Growth/Practice | 313 | 199 | 0.978 | 0.056 | 0.062 | Yes |
| Core three factors, 11 items | 313 | 41 | 0.996 | 0.030 | 0.040 | Yes |
| Core, merged Growth/Process | 313 | 43 | 0.974 | 0.073 | 0.065 | Yes |
| Core, merged Process/Gains | 313 | 43 | 0.981 | 0.064 | 0.056 | Yes |

### 4.2 Patterns of GenAI use

GenAI appears to have become a routine study tool for much of this sample (Figure 1): 85.6% used it “often” or “very frequently”, and 40.3% used it to assist with more than half of their academic tasks. Three to five interactions per assignment was the most widely selected response (36.4%), although a majority reported six or more (57.5%). Brief exchanges with GenAI could reflect simple questions, efficient prompting, or early acceptance of an answer with little further exploration. The last possibility raises a concern about superficial engagement, but interaction counts alone cannot distinguish these explanations. Research on authentic student dialogues similarly distinguishes participation from actively challenging and directing GenAI responses (Luo & Yin, 2026). The educational question is therefore what students do within and after these exchanges, rather than whether they simply make more requests.

### 4.3 Perceived skill development and changes in assignment processes

Students reported stronger improvement in efficient task execution than in evaluative judgment (Figure 2(a)). Endorsement was highest for improved efficiency (89.1%), followed by effective prompting (83.1%) and handling complex tasks (78.6%), while fewer students reported improvement in evaluating output accuracy (74.4%). The mean difference between efficiency and evaluation was 0.335 points ($d_z = 0.413$, with Holm-adjusted $p < 0.001$) (Table 5). This pattern suggests that greater perceived productivity with GenAI does not necessarily coincide with equivalent growth in evaluating its outputs. It is consistent with distinctions between application and evaluation in AI-literacy frameworks and with reports of lower confidence in evaluation and creation skills (Ng et al., 2021; O'Dea et al., 2026), although the present measures reflect perceived growth rather than absolute competence.

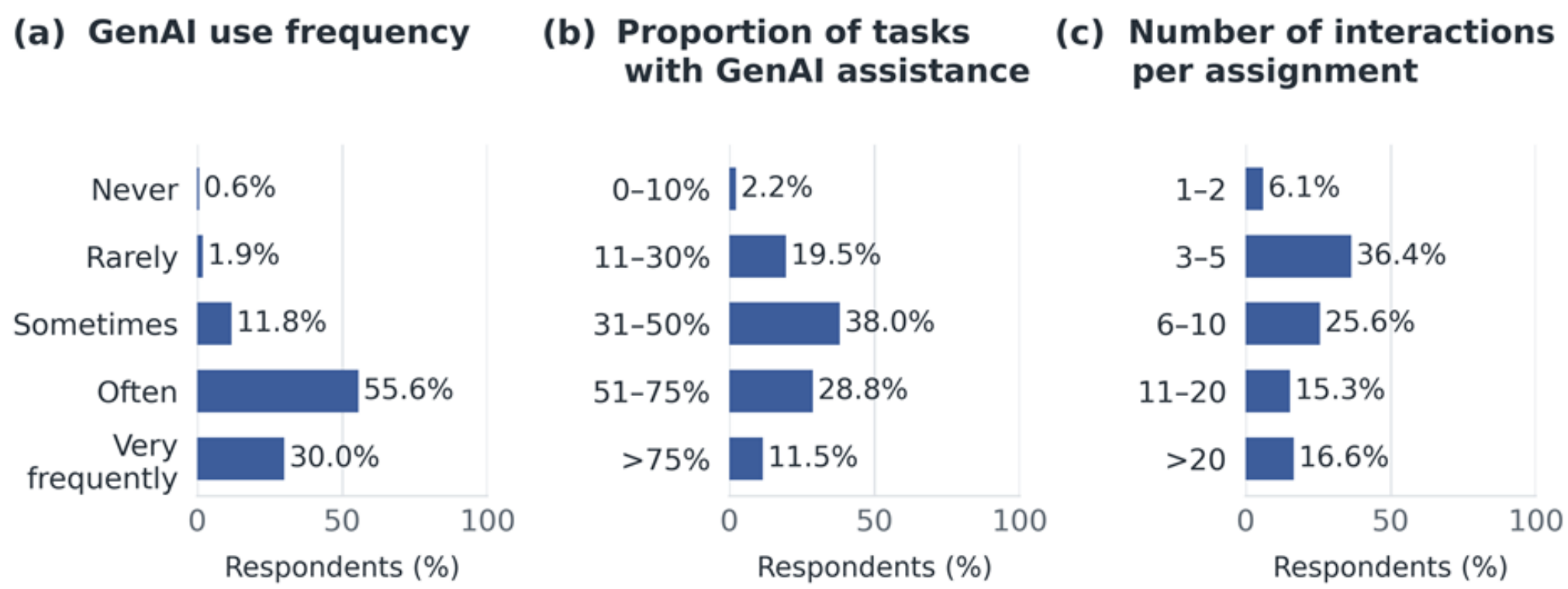

Figure 1. Patterns of GenAI use: (a) use frequency (H1), (b) proportion of academic assignment tasks completed with GenAI assistance (H2), and (c) interactions or prompts per assignment (H3). Percentages use $N$ = 313 and are displayed to one decimal place.

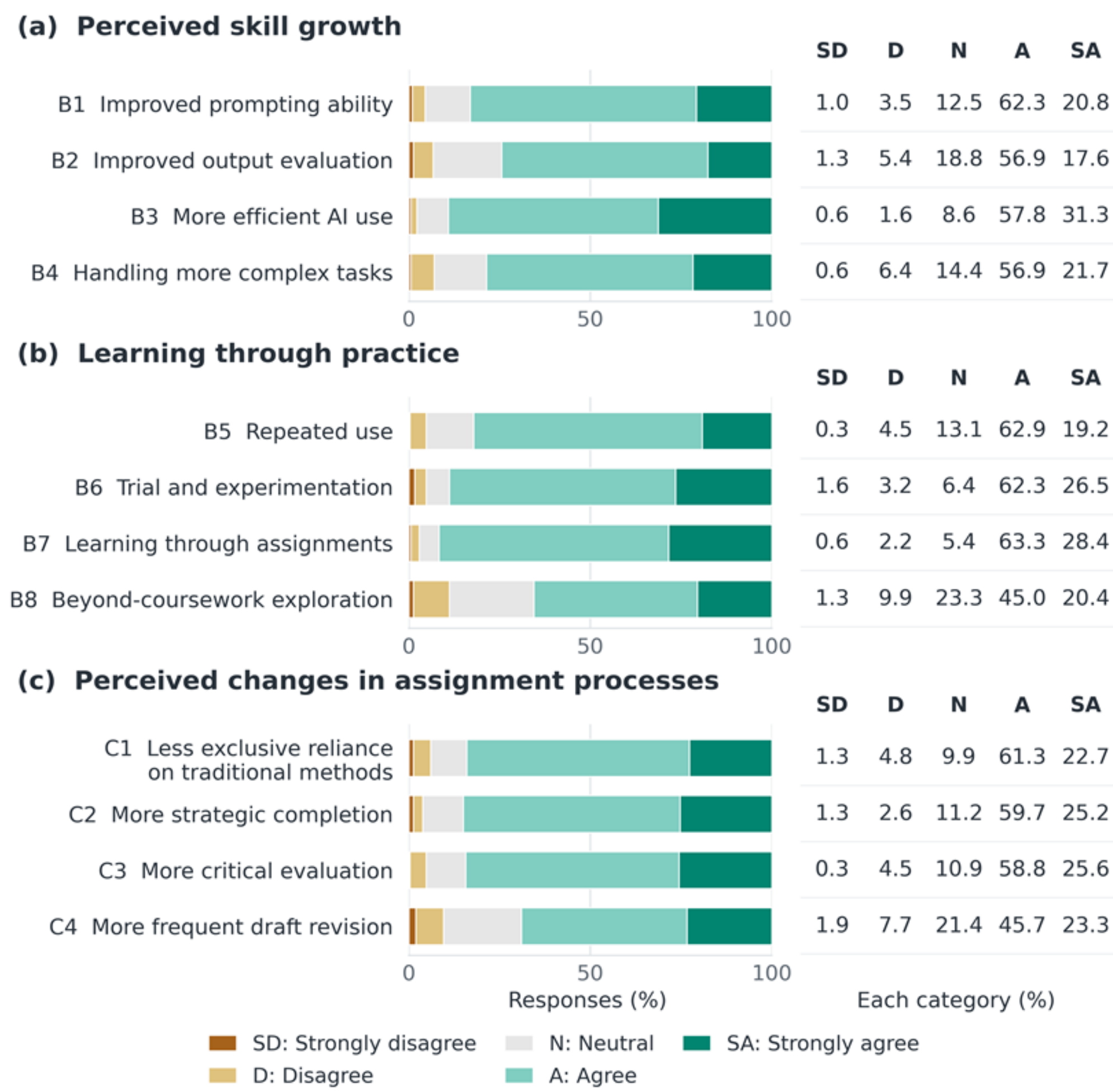

**Figure 2.** Response distributions for (a) perceived skill growth, (b) learning through practice, and (c) assignment-process adaptation.

Assignment-related learning was endorsed more frequently than learning new GenAI uses beyond coursework (91.7% vs. 65.4%; Figure 2(b)), suggesting that students' development was driven more by immediate academic demands than by broader exploration. A similar pattern appeared in assignment processes: strategic working (84.9%) and critical evaluation (84.4%) were endorsed more often than increased draft revision (69.0%; Figure 2(c)). Thus, students' perceived development was stronger in planning and monitoring aspects of GenAI use than in iterative revision. This pattern is consistent with self-regulated learning perspectives, which emphasize planning, monitoring, and revision as distinct components of learning regulation (Panadero, 2017), although self-reported endorsement does not establish the quality of these processes.

Interview accounts further illustrated how such development occurred. P05 described moving from using GenAI for paper summaries to asking targeted questions about research problems, experimental support, and project applications, while P11 attributed broader use to changing personal needs. P14 similarly used GenAI selectively for information structuring while retaining control over substantive content. Together, these accounts suggest a shift toward more purposeful and selective allocation of GenAI assistance as academic tasks and needs evolved, without demonstrating corresponding improvements in the quality of students' work.

Table 5. Exploratory within-person contrasts. Differences are first item minus second on the 1–5 scale. Bootstrap CI used 5,000 respondent resamples. All six Wilcoxon tests have Holm-adjusted $p < 0.001$.

| **Comparison** | **Mean difference** | **95% bootstrap CI** | $d_z$ |
|---|---|---|---|
| Efficiency − evaluative growth | 0.335 | [0.246, 0.425] | 0.413 |
| Assignment − examination gains | 0.204 | [0.128, 0.281] | 0.294 |
| Online − university support | 0.316 | [0.201, 0.435] | 0.302 |
| Peer − university support | 0.236 | [0.134, 0.345] | 0.248 |
| Assessment adaptation − reflection | 0.383 | [0.288, 0.476] | 0.450 |
| Assessment adaptation − appropriateness | 0.463 | [0.374, 0.556] | 0.554 |

### 4.4 Perceived academic gains and task-specific improvement

Perceived academic benefits varied across assessment contexts, with stronger gains reported for assignments than examinations (Figure 3(a)). Overall academic improvement was endorsed by 78.9% of students, while 74.7% reported assignment-grade gains compared with 61.4% for examination grades. The assignment–examination mean difference was 0.204 points ($d_z = 0.294$; Holm-adjusted $p < 0.001$; Table 5). This pattern suggests that benefits experienced during GenAI-supported assignment work may not transfer equally to examination settings.

The most frequently reported areas of improvement were conceptual understanding (65.8%), academic writing (63.6%), and brainstorming (61.7%) (Figure 3(b)), indicating that students associated GenAI use with both developing ideas and expressing them. Programming (27.8%) and citation work (28.1%) were less frequently selected, although these differences may reflect variation in task exposure as well as perceived benefit and should not be interpreted as rankings of GenAI effectiveness across tasks.

The interviews further clarified why these benefits may depend on assessment context. P01 attributed limited examination benefits partly to restrictions on GenAI use, whereas P13 described learning gains during examination preparation. P18 noted that better language did not necessarily translate into higher marks. These accounts distinguish improvements in GenAI-assisted work from gains in underlying understanding or assessed performance. This distinction is consistent with experimental evidence that improvements in assisted performance do not necessarily produce corresponding gains in knowledge, transfer, or subsequent unaided performance (Bastani et al., 2025; Y. Fan et al., 2025).

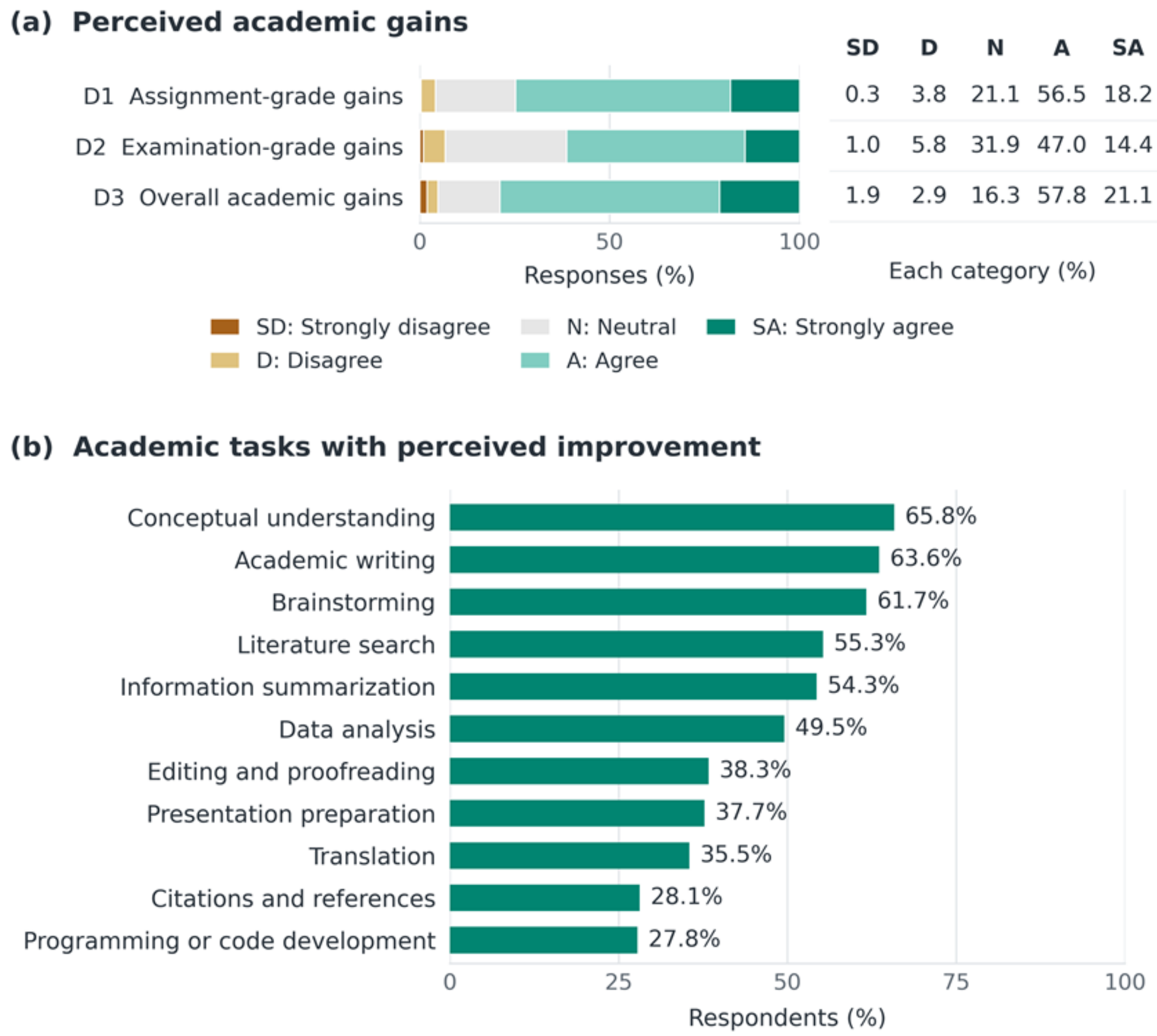

Figure 3. Perceived improvement: (a) perceived academic gains, (b) perceived task-specific improvement.

### 4.5 Retrospectively perceived developmental trajectories

Students reported several distinct patterns of perceived development rather than a single common trajectory (Figure 4(a)). The most frequent was rapid early improvement followed by a plateau (34.2%), followed by steady improvement (28.4%) and delayed acceleration (21.1%). Fluctuating development was reported by 12.8%, while little or no improvement (2.9%) and other patterns (0.6%) were uncommon. The prominence of plateau and delayed-acceleration patterns suggests that students did not generally perceive development as progressing at a constant rate.

Trajectory tests demonstrate that trajectory groups differed in several aspects of perceived development (Table 6). After Benjamini–Hochberg correction, significant overall differences were found for perceived skill growth, practice-related development, assignment-process adaptation, and responsible academic use, with the largest group difference observed for responsible academic use ($\varepsilon^2 = 0.073$). In contrast, perceived academic gains and assessment alignment showed no clear differences across trajectories, nor did time since first use or use frequency. Consistent with this, the three largest trajectory groups had similar mean academic-gain scores (3.82–3.91; Figure 4(b)). These findings indicate that perceived developmental trajectories were more closely differentiated by how students developed and regulated their GenAI use than by the academic benefits they reported. Accordingly, trajectory labels should not be interpreted as a simple ranking of academic benefit.

The interviews further illustrated why perceived trajectories may differ. P07 described little change in the purposes of GenAI use, whereas P08 reported broadly stable uses accompanied by greater fluency and more precise prompting. Other participants linked turning points to changing academic demands or improvements in the technology itself; for example, P13 and P18 emphasized advances in GenAI capabilities. These accounts suggest that students attributed perceived development to an interplay of evolving personal strategies, changing task demands, and technological change rather than to accumulated experience alone. This interpretation is consistent with evidence of uneven development in academic GenAI practices reported elsewhere (Aruleba et al., 2025).

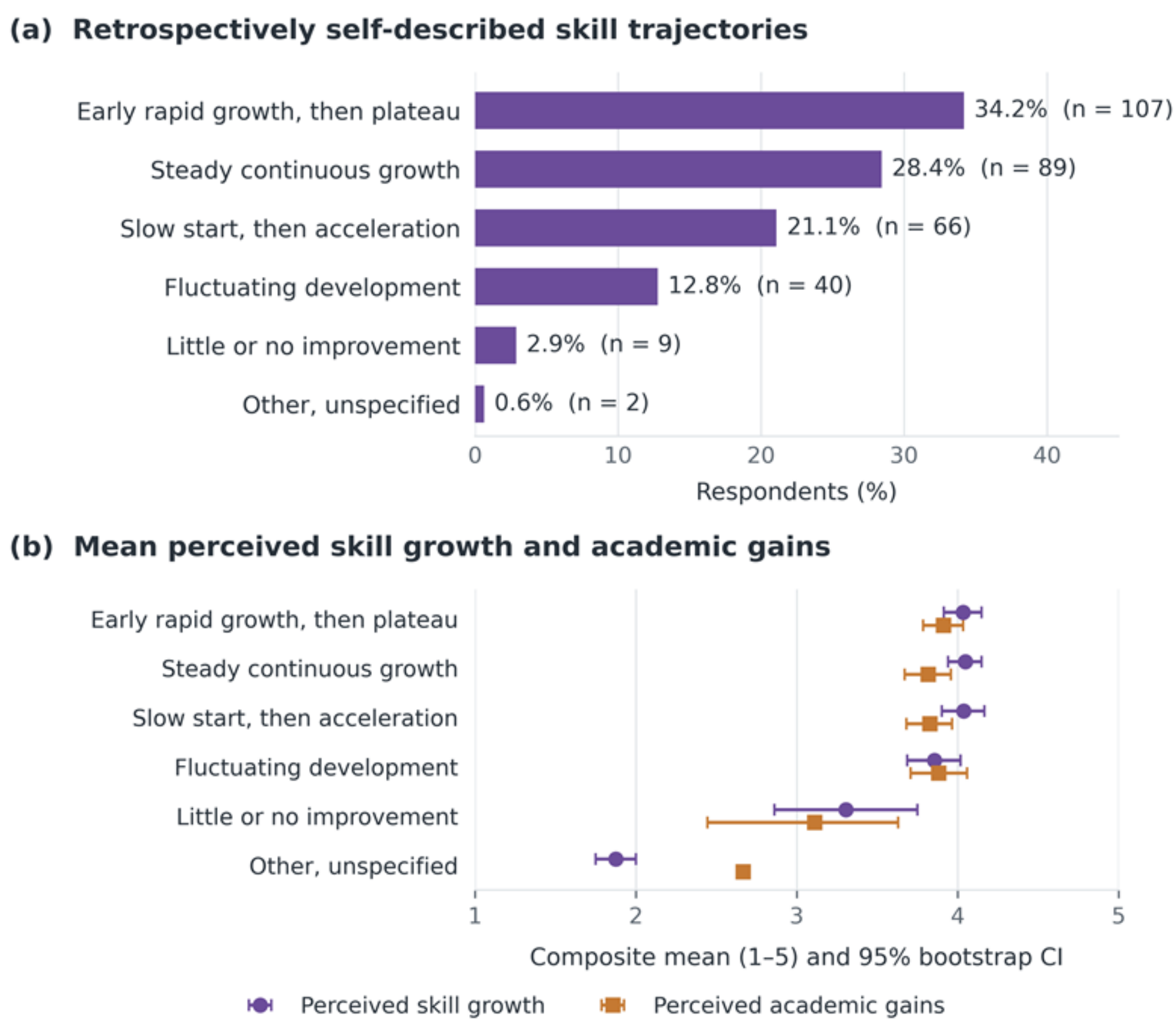

Figure 4. Retrospectively perceived developmental trajectories and group-level scores. (a) All six nominal response categories. (b) Means of the growth and academic-gains composites on their 1–5 scales, with pointwise percentile intervals from 5,000 within-group respondent resamples. The same resampled respondents were used for both composites.

Table 6. Trajectory-group comparisons based on 311 of the 313 respondents (two unspecified "other" responses were excluded from these analyses)

| **Variable** | $H(4)$ | $p$ | **BH-adjusted** $p$ | $\varepsilon^2$ |
|---|---|---|---|---|
| Perceived growth | 15.261 | 0.004 | 0.011 | 0.037 |
| Practice-related development | 18.842 | < 0.001 | 0.003 | 0.049 |
| Process adaptation | 14.448 | 0.006 | 0.012 | 0.034 |
| Perceived academic gains | 7.142 | 0.129 | 0.185 | 0.010 |
| Assessment alignment | 6.949 | 0.139 | 0.185 | 0.010 |
| Responsible use | 26.305 | < 0.001 | < 0.001 | 0.073 |
| Time since first use | 0.797 | 0.939 | 0.939 | 0.000 |
| Use frequency | 3.404 | 0.493 | 0.563 | 0.000 |

### 4.6 Sources of support and assessment alignment

Course requirements appeared to stimulate GenAI development, but students also relied heavily on support beyond formal university teaching (Figure 5(a)). Course requirements were endorsed by 79.8% of students, online resources by 78.0%, peer exchange by 72.2%, and university teaching or workshops by 61.0%. Within-person comparisons also showed higher ratings for online resources and peer support than for university teaching, with mean differences of 0.316 and 0.236 points, respectively (both Holm-adjusted $p < 0.001$; Table 5). Together with their higher endorsement rates, these differences suggest that students drew more extensively on online and peer-based support. The comparatively lower rating of university support points to potential gaps in its reach, uptake, or relevance to students' needs, rather than necessarily indicating lower instructional effectiveness.

The interviews help explain this pattern by distinguishing access to support from participation and perceived relevance. Several participants described university provision that they had not attended, while P12 found formal teaching valuable for discovering and combining GenAI tools. Others questioned its relevance: P08 recalled image-generation instruction that had little application to their work, while P14 described some material as already familiar and other content as too advanced. Thus, lower ratings of university support may reflect uneven exposure and fit with students' needs rather than limited instructional value.

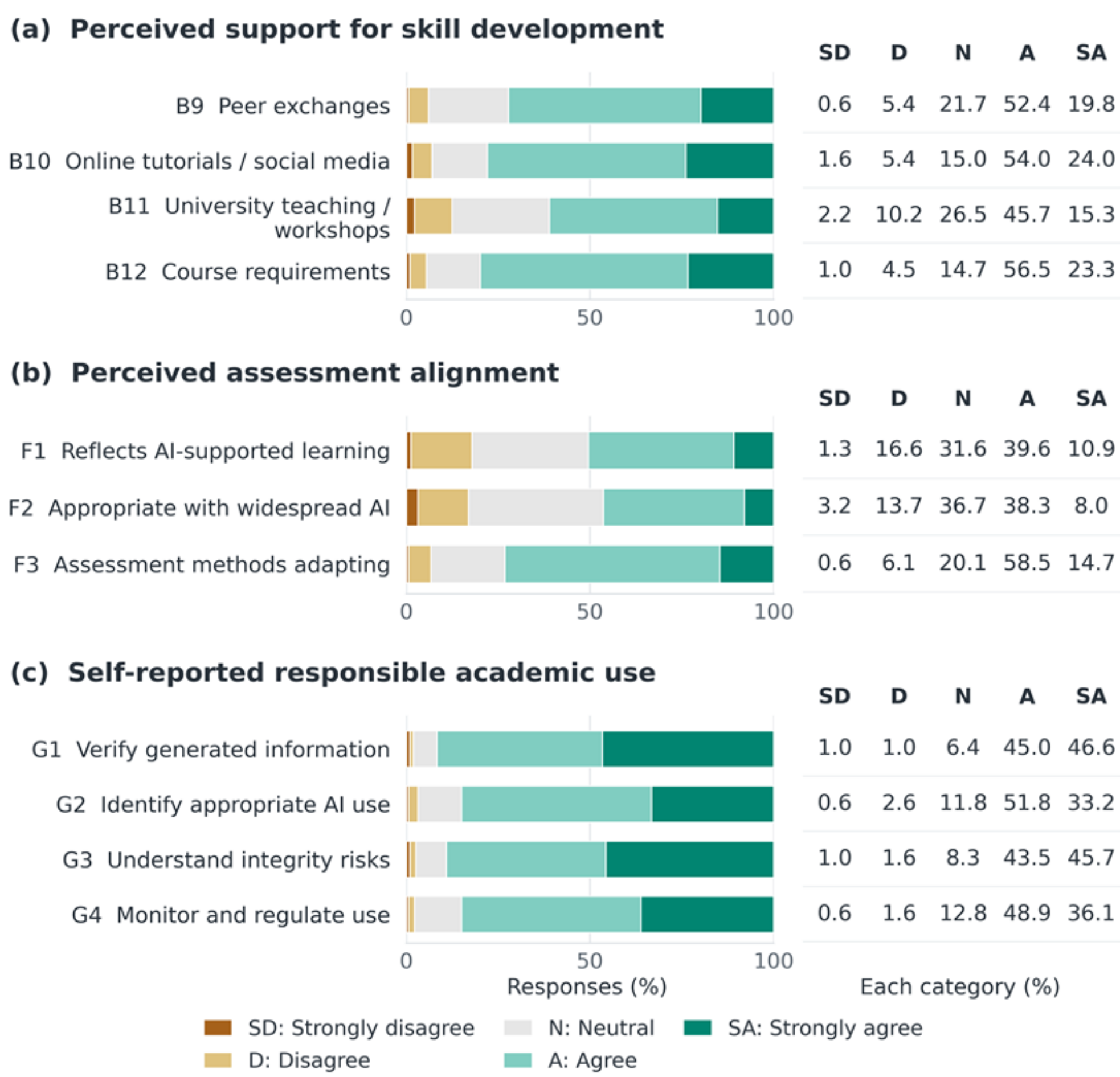

Figure 5. Response distributions for (a) learning support sources, (b) assessment alignment, and (c) perceived responsible GenAI use (self-reported behavior rather than demonstrated competence or observed compliance).

A similar gap emerged in perceptions of assessment alignment (Figure 5(b)). Although 73.2% perceived that assessment had adapted to GenAI, only 46.3% considered current assessment suitable for widespread GenAI use. The mean difference was 0.463 points ($dz = 0.554$; Holm-adjusted $p < 0.001$; Table 5). This suggests that visible adaptation has not necessarily translated into confidence that assessment adequately reflects GenAI-supported learning. The substantial neutral responses further indicate uncertainty about this alignment. Constructive alignment and the AI Assessment Scale provide useful frameworks for considering how assessment can clarify intended learning, students' own contribution, and permissible AI assistance (Biggs, 1996; Perkins et al., 2024).

### 4.7 Responsible use

Students generally reported high levels of responsible GenAI use (Figure 5(c)). Verification was endorsed by 91.6%, awareness of academic-integrity risks by 89.2%, and both identifying appropriate uses and monitoring or regulating use by 85.0%. However, this strong endorsement of verification contrasted with the lower reported growth in evaluating output accuracy, suggesting that performing checks is not necessarily the same as becoming more capable of judging whether an output is correct.

The interviews showed that verification took several forms. Some relied on direct checking against external evidence. After encountering fabricated references, P03 developed the habit: "I ask it to send me the link to the paper, and then check again whether the paper actually exists". P07 manually checked calculations and counts after identifying errors, P16 returned to original sources to verify reported outputs, and P17 identified issues by running code and manually checking source material. Others used comparative approaches: P09 compared outputs across tools and adapted prose to a preferred style, while P14 treated agreement between different GenAI responses as a form of checking. These accounts demonstrate varied verification practices behind the survey findings, although such practices do not necessarily ensure accurate judgments.

The interviews also highlighted a distinction between checking outputs and having confidence in evaluating them. P18 acknowledged this uncertainty: "Not really; my judgment is not yet that strong". Despite limited confidence, P18 described consulting authoritative sources for data and citations, suggesting that verification behaviours can coexist with uncertainty about evaluation quality.

P15 further emphasized the role of disciplinary knowledge. After initially accepting an inaccurate answer on an unfamiliar tax-law topic, P15 concluded that judging GenAI outputs depended on understanding the relevant field. Thus, verification strategies alone may be insufficient when users lack the knowledge required to evaluate evidence effectively. Together, these findings suggest that effective GenAI use requires not only checking behaviours but also informed evaluation capabilities, consistent with AI-literacy frameworks emphasizing critical assessment alongside application (Ng et al., 2021).

### 4.8 Recommendations for institutional support

The open-response item yielded 83 valid comments, with recommendations centering on practical guidance and preserving students' independent engagement with academic work. Training or guidance was the most common theme (49.4%), followed by independent thinking (39.8%), rules or academic integrity (19.3%), assessment redesign (12.0%), and access to GenAI tools or resources (10.8%) (Table 7). Because individual comments could receive multiple codes, these percentages describe the thematic composition of the 83 responses rather than prevalence across the full sample.

The comments also revealed how students connected responsible GenAI use with assessment design and academic ownership. One respondent proposed, "Include oral defenses in assessment so students truly understand their own or GenAI-generated work", while another urged educators to "strengthen assignment design through critical thinking, practical work, and originality". Views on regulation were less consistent: some respondents called for stronger detection of GenAI-generated content, whereas others objected to monitoring the proportion of GenAI-generated text in student work. Overall, the comments indicate broad concern with supporting meaningful and responsible GenAI use, but no clear consensus on the institutional policies best suited to achieve it.

Table 7. Content codes in institutional-support suggestions. One comment could receive multiple codes.

| **Content code** | ***n*** | **%** |
|---|---|---|
| Providing practical instruction and guidance on GenAI use | 41 | 49.4 |
| Promoting independent thinking and avoiding overreliance | 33 | 39.8 |
| Establishing policies on academic integrity, disclosure, and detection | 16 | 19.3 |
| Redesigning assignments and assessment for learning with GenAI | 10 | 12.0 |
| Providing access to GenAI tools and resources | 9 | 10.8 |
| Developing students' ability to evaluate and verify GenAI outputs | 9 | 10.8 |
| Integrating GenAI into teaching | 7 | 8.4 |

### 4.9 Delegation and dependence

Interview accounts suggest that the educational value of strategic GenAI use depends partly on which cognitive steps students continue to perform themselves. P05 described thinking through a problem before consulting GenAI and requesting guided steps rather than an immediate answer, while P18 similarly reported thinking first and seeking help only when stuck. By contrast, P08 and P11 associated deadline pressure with reduced independent thinking or greater reliance on GenAI. P16 described bypassing intermediate reasoning altogether: "I may skip those steps directly, so there is less of my own logical thinking". These accounts distinguish using GenAI to support reasoning from using it to replace parts of that process.

Productive assistance and dependence could also coexist within the same student's account. P12 described learning through stepwise questioning while also reporting increased laziness. Similarly, P13 rejected the idea that skilled GenAI use was inherently lazy but acknowledged that following an GenAI-generated plan could reduce consideration of the problem as a whole. Dependence therefore appeared neither uniform nor fixed, but contingent on the task, circumstances, and way assistance was used.

These tensions qualify the survey's generally favorable ratings of process adaptation and responsible use. Students may organize GenAI assistance effectively while still relinquishing opportunities for independent reasoning. Similar tensions have been reported elsewhere, including the coexistence of improved prompting with persistent copying and contrasting relationships between GenAI use, self-efficacy, and thinking offloading (Aruleba et al., 2025; Ma et al., 2026). The interviews do not demonstrate cognitive decline or a causal effect of GenAI use on dependence; rather, they suggest that the educational consequences of delegation depend on which elements of reasoning students retain.

Delegation could also have a facilitative dimension. P18 reported that GenAI reduced embarrassment about asking questions and allowed assistance to be sought without the interpersonal pressure associated with approaching teachers. Thus, GenAI may lower barriers to support while simultaneously raising questions about how much reasoning students delegate.

### 4.10 Exploratory structural and contextual associations

The focused three-factor ordinal model demonstrated close approximate fit (Table 4), with an admissible latent correlation matrix. Within this model, perceived skill growth was strongly associated with perceived assignment-process adaptation ($\beta$ = 0.729), and perceived assignment-process adaptation was positively associated with perceived academic gains after accounting for perceived skill growth ($\beta$ = 0.620). In contrast, the direct association between perceived skill growth and academic gains was smaller and statistically uncertain ($\beta$ = 0.209, robust-test $p$ = 0.156).

The indirect association between perceived skill growth and academic gains through assignment-process adaptation was positive ($\beta$ = 0.452), resulting in a total association of $\beta$ = 0.661 (Table 8; Figure 6). This pattern suggests that students' perceived benefits may be linked not only to developing GenAI-related skills but also to how those skills are integrated into assignment-related processes, including strategy adjustment and reflective practice. This interpretation is consistent with self-regulated learning perspectives that emphasize monitoring and adaptation during learning activities (Panadero, 2017).

However, because the structural component among the three latent constructs was saturated, the reported fit indices primarily reflect the measurement restrictions rather than confirming the proposed directional relationships. The model-implied $R^2$ values were 0.531 for perceived assignment-process adaptation and 0.616 for perceived academic gains within the analytical sample, indicating the proportion of variance represented by the specified model rather than causal explanatory power or predictive performance. Furthermore, the positive indirect association and the imprecise direct association do not establish complete mediation. Given the cross-sectional design, the analysis cannot determine temporal ordering, demonstrate causal mechanisms, or exclude alternative explanations and unmeasured confounding.

Table 8. Standardized associations in the exploratory core ordinal model. Growth = perceived skill growth; Process = perceived assignment-process adaptation; Gains = perceived academic gains. The symbol "|" indicates statistical adjustment for the variable following it.

| **Association** | **Standardized *β*** | **95% bootstrap CI** | **Robust-test *p*-value** |
|---|---|---|---|
| Growth → Process | 0.729 | [0.601, 0.866] | < 0.001 |
| Process → Gains \| Growth | 0.620 | [0.328, 0.958] | < 0.001 |
| Growth → Gains \| Process | 0.209 | [-0.143, 0.510] | 0.156 |
| Indirect association via Process | 0.452 | [0.236, 0.763] | < 0.001 |
| Total Growth–Gains association | 0.661 | [0.515, 0.801] | < 0.001 |

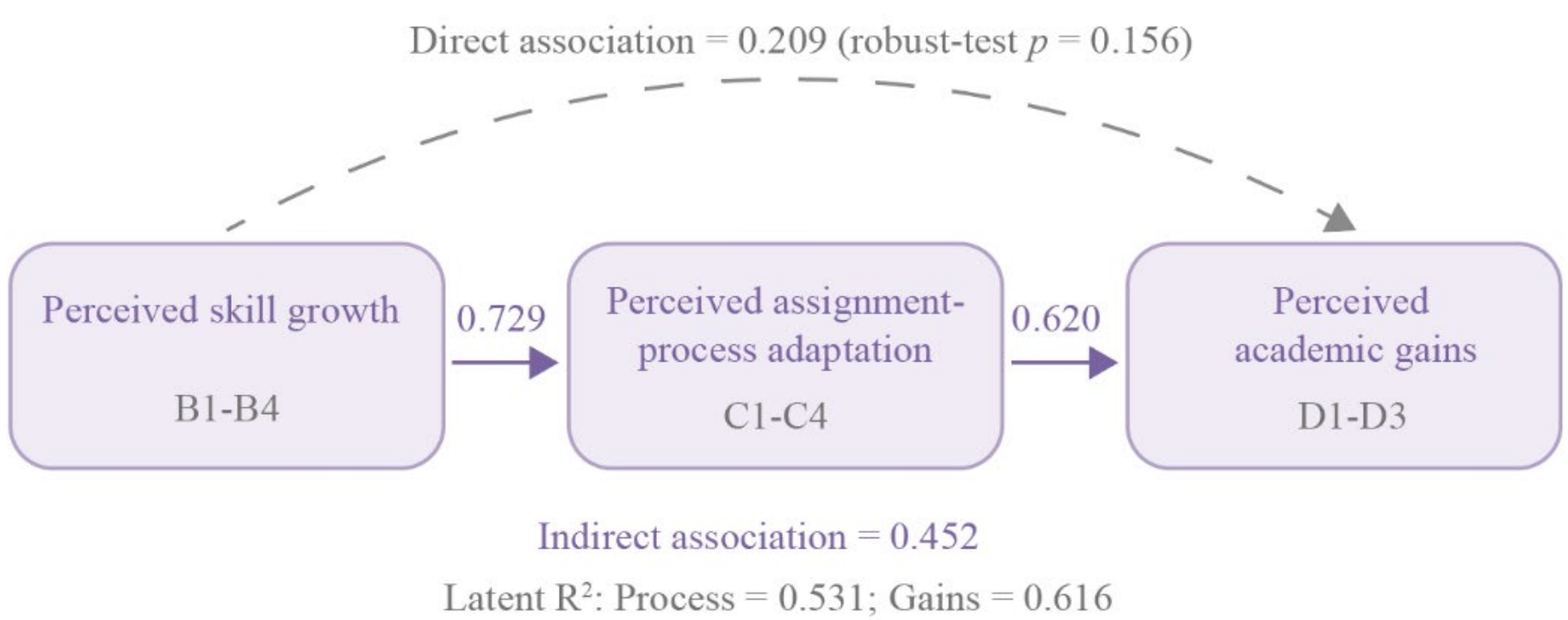

Figure 6. Exploratory three-factor ordinal model of perceived skill growth, perceived assignment-process adaptation, and perceived academic gains. Values on the arrows are fully standardized path coefficients Arrows parameterize concurrent associations and do not establish temporal ordering or causation. The dashed direct path has a 95% case-bootstrap CI that includes zero. $R^2$ denotes the proportion of latent outcome variance represented by the predictors under the specified model; Process = perceived assignment-process adaptation; Gains = perceived academic gains.

Contextual correlations provided a complementary, bivariate perspective (Figure 7). Whereas Figure 6 reports standardized associations among latent constructs, Figure 7 reports Spearman correlations involving survey composite scores. Perceived academic gains were positively associated with GenAI use frequency ($\rho$ = 0.216, BH-adjusted $p < 0.001$) and interactions per assignment ($\rho$ = 0.159, BH-adjusted $p$ = 0.008), but not clearly with the proportion of assignment tasks receiving GenAI assistance ($\rho$ = 0.081, BH-adjusted $p$ = 0.180) or time since first academic GenAI use ($\rho$ = 0.108, BH-adjusted $p$ = 0.072). Thus, different indicators of GenAI exposure and intensity showed different relationships with perceived academic gains rather than a uniform pattern.

Self-rated digital literacy was positively associated with perceived skill growth ($\rho$ = 0.419), assignment-process adaptation ($\rho$ = 0.289), perceived academic gains ($\rho$ = 0.284), and perceived responsible academic use ($\rho$ = 0.383), all BH-adjusted $p < 0.001$. Self-rated academic performance relative to peers showed a similar, though uneven, pattern, with the strongest correlation observed for responsible academic use ($\rho$ = 0.409). Overall, contextual associations varied across indicators and outcomes. As bivariate correlations, they do not establish independent effects or imply that increasing digital literacy or GenAI use would cause improvements in academic processes or outcomes.

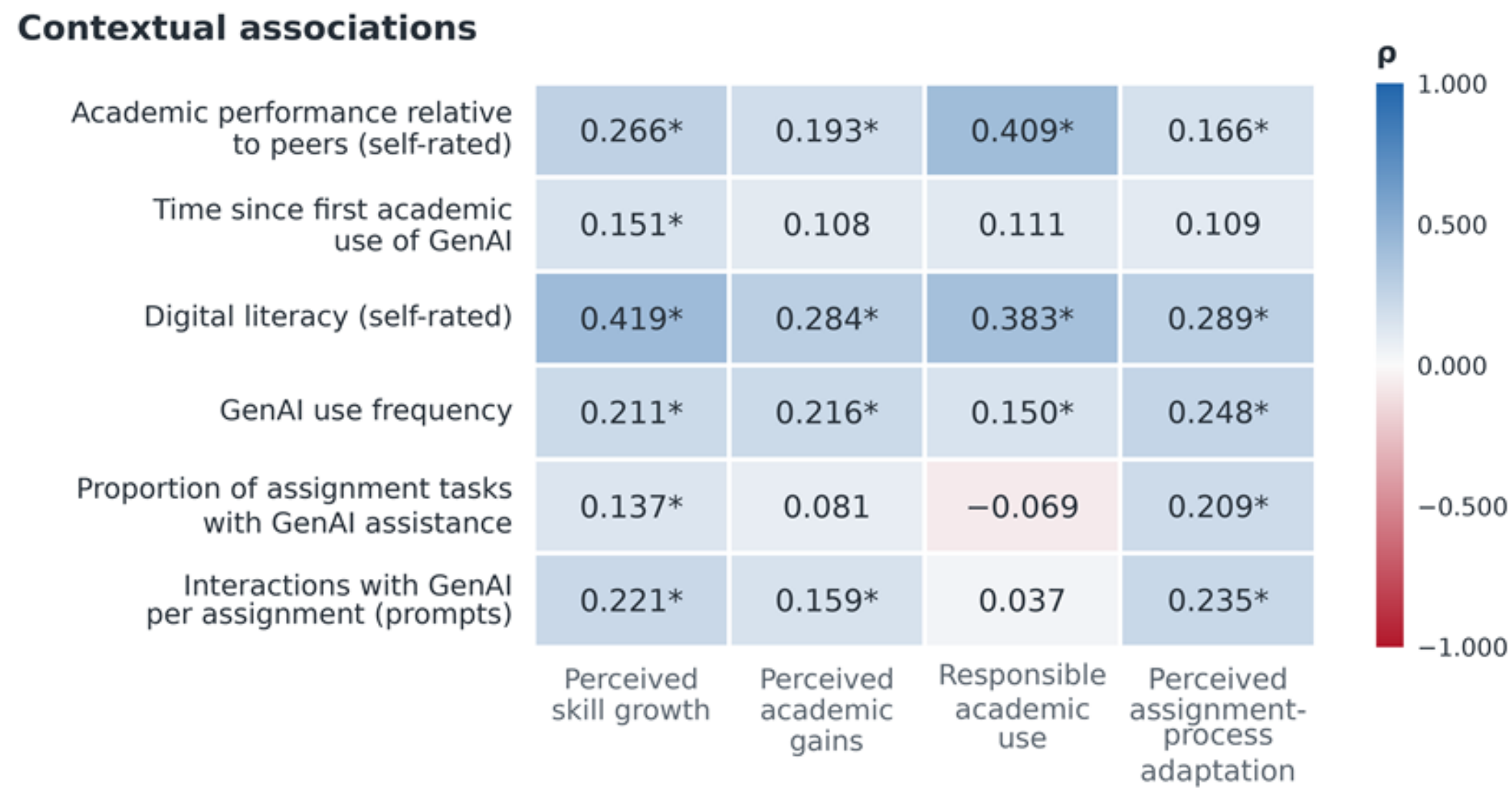

Figure 7. Spearman correlations ($\rho$) between six contextual measures (A4, A5, A7, H1–H3) and four student-focused survey composites means ($N = 313$): perceived skill growth, perceived academic gains, responsible academic use, and perceived assignment-process adaptation. The common color scale ranges from −1 to +1; an asterisk marks indicates BH-adjusted $p < 0.05$ across the complete 66-test correlation family. These are bivariate associations, not as sociations adjusted for other variables.

## 5. Discussion

### 5.1. Perceived heterogeneous and uneven development over time

Students did not perceive GenAI development as a uniform process. Interview accounts help explain this variation: students described changing academic tasks, increasingly specific requests, stable purposes of use, and improvements in GenAI tools themselves. Similar situated studies have linked evolving GenAI practices to changing academic demands and contexts (Aruleba et al., 2025; Tan, 2025).

Development was also uneven across capabilities. Perceived improvement in efficiency was more widely endorsed than improvement in evaluating output accuracy (89.1% vs. 74.4%). This distinction is consistent with AI-literacy frameworks that separate application from evaluation (Ng et al., 2021), and with evidence of lower confidence in evaluation and creation skills (O'Dea et al., 2026). Although confidence and perceived growth are not equivalent constructs, both point to the need to distinguish becoming more efficient with GenAI from becoming better able to judge its outputs.

Importantly, developmental trajectory and perceived academic benefit were not closely aligned. Although several developmental measures differed across trajectory groups, perceived academic gains did not clearly differ after correction. This finding does not establish equivalence, particularly given the small sizes of some groups, but it does indicate that developmental pattern and academic benefit should not be treated as interchangeable. A reported plateau, for example, could reflect stable routines, increasing task difficulty, changing standards, or limited feedback rather than an absence of learning. Repeated observations on comparable tasks, ideally under stable tool conditions, would be needed to distinguish learner development from these alternative explanations.

### 5.2. Learning processes and perceived performance gains

Perceived academic benefits also varied by assessment context. Assignment-grade gains were reported more frequently than examination-grade gains (74.7% vs. 61.4%). Interview accounts suggested several reasons for this difference, including restrictions on GenAI use in examinations and uncertainty about whether improved work quality translates into higher marks. This distinction is consistent with experimental evidence showing stronger gains in assisted performance than in knowledge gain or transfer (Y. Fan et al., 2025). Although the present study did not directly measure transfer, its findings caution against treating better GenAI-supported work as evidence of stronger independent learning.

The structural results reinforce the importance of how GenAI is used, not only whether skills are perceived to improve. The positive indirect association through assignment-process adaptation ($\beta = 0.452$) suggests that changes in academic practices may be relevant to the relationship between perceived skill growth and academic gains. Self-regulated learning provides a useful interpretive framework because it emphasizes strategy selection, monitoring, and revision (Panadero, 2017). However, the present process measures capture only part of this framework, and their concurrent, attributional nature does not establish a developmental sequence from skill growth to academic achievement.

The interviews further showed that delegation to GenAI could either support or bypass learning. P12 described both stepwise learning and increasing reliance, whereas P16 reported skipping intermediate reasoning. This tension resembles Ma et al.'s (2026) contrasting associations involving self-efficacy and thinking offloading. Intervention studies similarly suggest that outcomes depend on how GenAI support is designed and used (Bastani et al., 2025; Kestin et al., 2025). The central issue is therefore not simply the extent of GenAI use, but which cognitive responsibilities students retain for explanation, verification, and judgment.

### 5.3. Assessment adaptation

Assessment findings reveal a related gap between visible change and perceived adequacy. While 73.2% of students perceived that assessment practice had adapted to GenAI, only 46.3% considered current assessment appropriate for widespread GenAI use. From a constructive-alignment perspective (Biggs, 1996), changing task formats or GenAI permissions is insufficient unless assessment continues to capture the intended learning outcomes. The gap between adaptation and appropriateness therefore points to the need for assessment designs that make students' reasoning and contribution visible rather than evaluating only the final product.

Frameworks such as Chan's policy model and the AI Assessment Scale offer ways to clarify expectations and define permissible levels of assistance (Chan, 2023; Perkins et al., 2024). Practices such as draft rationales, oral explanations, or comparisons between generated and independently verified answers may help reveal how students reason with GenAI. Their effectiveness, feasibility, and workload implications, however, should be tested within specific disciplinary and assessment contexts rather than assumed.

### 5.4. Limitations and future research

Several limitations should be considered. Both survey and interview data were retrospective self-reports and therefore capture perceived rather than directly observed development. The study did not measure individual growth rates, verified grades, or unaided competence, and responses may have been influenced by recall bias, social desirability, response tendencies, and attributional item wording. Overlap between some interview questions and survey constructs also limits the independence of the qualitative evidence, while factor analysis and one-factor model comparisons cannot rule out common-method bias (Podsakoff et al., 2003).

The cross-sectional design precludes conclusions about temporal ordering or causality. Associations among perceived growth, process adaptation, and academic gains do not show that one preceded or caused another, and the indirect association should not be interpreted as mediation. Retrospective trajectories may also reflect changes in students, academic demands, or GenAI capabilities, alongside unmeasured factors such as motivation, prior ability, discipline, and GenAI experience.

Future research should consider prospective longitudinal and multimethod designs combining repeated competency assessments, GenAI interaction logs, assignment artefacts, verified academic outcomes, and linked interviews. Such studies should distinguish learner development from changes in tool capability, task demands, and disciplinary context. Importantly, comparisons between GenAI-assisted and unaided performance are needed to determine whether perceived gains translate into durable learning and transfer. Experimental or quasi-experimental studies could further test whether verification training, reflective use, appropriate delegation guidance, and assessment redesign improve both effective GenAI use and independent academic competence.

## 6. Conclusion

This study identified university students' self-perceived GenAI skill development as heterogeneous over time. Among 313 survey respondents, rapid early improvement followed by a plateau was the most common retrospective trajectory (34.2%), followed by steady improvement (28.4%) and delayed acceleration (21.1%). Development also varied across capabilities, with improvement in evaluating GenAI outputs perceived as more challenging than improvement in other academic skills.

Perceived development was accompanied by changes in learning practices. Students who reported greater skill growth also tended to report greater adaptation of assignment processes, while assignments were the most frequently selected driver of skill development. Interviews further showed that perceived progress could involve more specific prompting, more varied forms of verification, and more selective delegation. However, some students described stable routines, reliance on GenAI that bypassed reasoning, or improvement attributable partly to advances in the tools themselves. These findings caution against treating greater efficiency alone as evidence of stronger independent learning.

Academic benefits were similarly differentiated. Assignment-grade gains were reported more often than examination-grade gains (74.7% vs. 61.4%), while perceived academic gains did not clearly differ across developmental trajectory groups. The structural analysis further indicated a positive indirect association between perceived skill growth and academic gains through adaptation of assignment practices, suggesting that how students integrate GenAI into academic work may be more consequential than the particular trajectory they perceive themselves to follow.

These findings highlight the importance of supporting students not only in using GenAI effectively but also in evaluating outputs, retaining responsibility for reasoning, and demonstrating understanding alongside GenAI-assisted performance. Interviewees also emphasized the need for more tailored training that addresses students' differing skill levels, disciplinary contexts, and specific learning needs. Such support, together with appropriately designed assessment practices, may better accommodate varied patterns of GenAI development and use. Because the study relies on retrospective self-reports collected at one point in time, it cannot establish actual longitudinal development or causal academic effects. Future research should therefore use repeated measures of GenAI competence, learning practices, and academic performance to determine whether these perceived developmental patterns correspond to durable changes in learning.

**Data availability**

All data supporting the findings of this study are available within the paper.

**Competing interest**

The authors declare no competing interests.

**Ethics approval**

The study protocols for the survey and the interview were approved by the University Research Ethics Review Panel of Xi'an Jiaotong-Liverpool University on 13 Jun 2026. All procedures were conducted in accordance with applicable ethical guidelines and regulations.

**Informed consent**

Informed consent was obtained from all individual participants for participation in this study. Anonymity of their identification and voluntary participation were assured. No identifiable data of participants were present within the manuscript.